\documentclass[aps,prd,reprint,groupedaddress]{revtex4-1}
\usepackage{graphicx}
\usepackage{amssymb, amsmath,mathtools,amsthm}
\newtheoremstyle{dotless}{}{}{\itshape}{}{\bfseries}{}{ }{}
\theoremstyle{dotless}
\makeatletter
\def\@endtheorem{\endtrivlist}% NEW
\makeatother
\newtheorem*{proposition*}{}

\newcommand{\be}{\begin{equation}}
\newcommand{\bea}{\begin{align}}
\newcommand{\eea}{\end{align}}
\newcommand{\beq}{\begin{equation}}
\newcommand{\ee}{\end{equation}}
\newcommand{\eeq}{\end{equation}}

\newcommand{\mnras}{{MNRAS}}		
		
\newcommand{\pasj}{{PASJ}}

\newcommand{\ssr}{{Space~Sci.~Rev.}}

\newcommand{\physscr}{{Phys.~Scr}}

\begin{document}

%\preprint{}

\title{The Black Hole Meissner Effect and  Blandford-Znajek Jets}


\author{Robert F. Penna}
\email[]{rpenna@mit.edu}
%\thanks{}
\affiliation{Department of Physics and Kavli Institute for Astrophysics and Space Research,
Massachusetts Institute of Technology, Cambridge, Massachusetts 02139, USA}


\date{\today}

\begin{abstract}

Spinning black holes tend to expel magnetic fields.  In this way they are similar to superconductors.  It has been a persistent concern that this black hole ``Meissner effect'' could quench jet power at high spins.  
This would make it impossible for the rapidly rotating black holes in Cyg X-1 and GRS 1915+105 to drive Blandford-Znajek jets.  We give a simple geometrical argument why fields which become entirely radial near the horizon are not expelled by the Meissner effect and may continue to power jets up to the extremal limit.  A simple and natural example is a split-monopole field.   We stress that ordinary Blandford-Znajek jets are impossible if the Meissner effect operates and expels the field.  Finally, we note that in our general relativistic magnetohydrodynamic simulations of black hole jets, there is no evidence that jets are quenched by the Meissner effect.  The simulated jets develop a large split monopole component spontaneously which supports our proposal for how the Meissner effect is evaded and jets from rapidly rotating black holes are powered in nature.

\end{abstract}

\pacs{}
%\keywords{}

\maketitle

\section{Introduction}
\label{sec:intro}

Spinning black holes tend to expel magnetic fields.  Astrophysical jets are believed to be powered by magnetized, spinning black holes.  Magnetic fields need to thread the horizon to extract the black hole's rotational energy (a point we will return to later).  So if the black hole Meissner effect prevents rapidly rotating black holes from becoming magnetized, it could quench jet power.  

Until recently, one might have argued that astrophysical black holes do not achieve the high spins where the Meissner effect is important.  However, there is now reliable evidence for rapidly rotating black holes.  In particular, the spin parameters of the black holes in Cyg X-1 and GRS 1915+105 have been measured to be $a/M>0.95$ \cite{2006ApJ...652..518M,2011ApJ...742...85G,2013SSRv..tmp...73M}.  Jets from these black holes could be quenched by the Meissner effect.   It is important to understand this possibility.

\begin{figure*}[!ht]
\begin{center}
\includegraphics[width=0.45\textwidth]{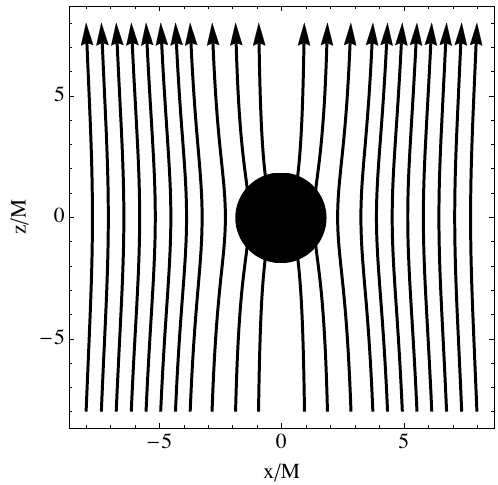}
\includegraphics[width=0.45\textwidth]{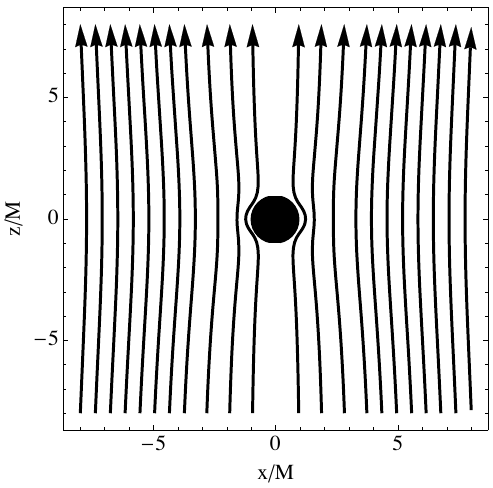}
\caption{The Wald magnetic field \cite{1974PhRvD..10.1680W} for black hole spin parameters $a/M=0.5$ (left panel) and $a/M=1$ (right panel).  At $a/M=1$, the field is completely expelled from the horizon.}
\label{fig:waldfields}
\end{center}
\end{figure*}

The discovery of the black hole Meissner effect predates astrophysical jet modeling.  Wald \cite{1974PhRvD..10.1680W} found a solution for a Kerr black hole immersed in a uniform magnetic field aligned with the black hole spin axis.  The magnetic field is treated as a test field.  It is a vacuum field; there are no currents.  The simplicity of Wald's solution makes it very useful for understanding the interaction of black holes with magnetic fields.  King, Lasota, and Kundt \cite{1975PhRvD..12.3037K} noted that the flux of Wald's solution through the black hole horizon drops to zero in the extremal limit (see Figure \ref{fig:waldfields}), in a way that is similar to the Meissner effect of superconductors  \cite{1980PhRvD..22.2933B,1998PhRvD..58h4009C}.

\begin{figure*}[!ht]
\begin{center}
\includegraphics[width=0.45\textwidth]{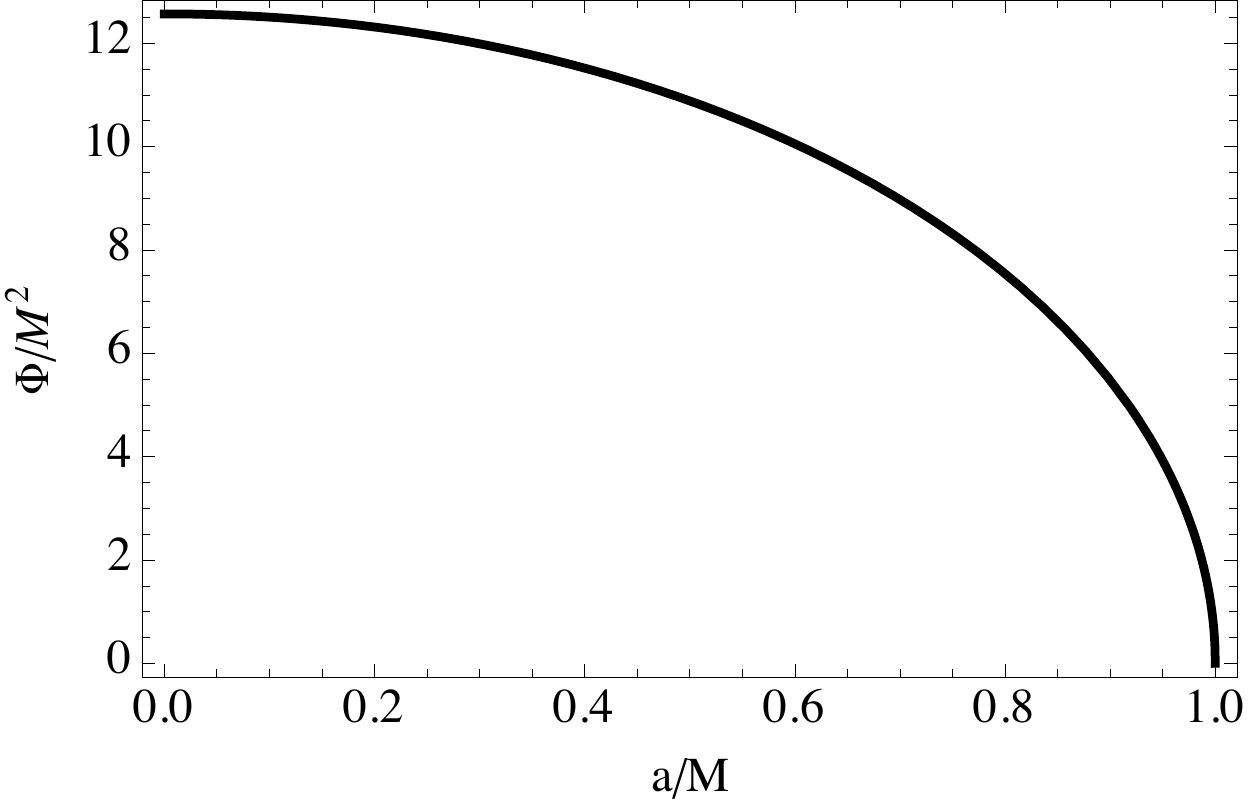}
\includegraphics[width=0.45\textwidth]{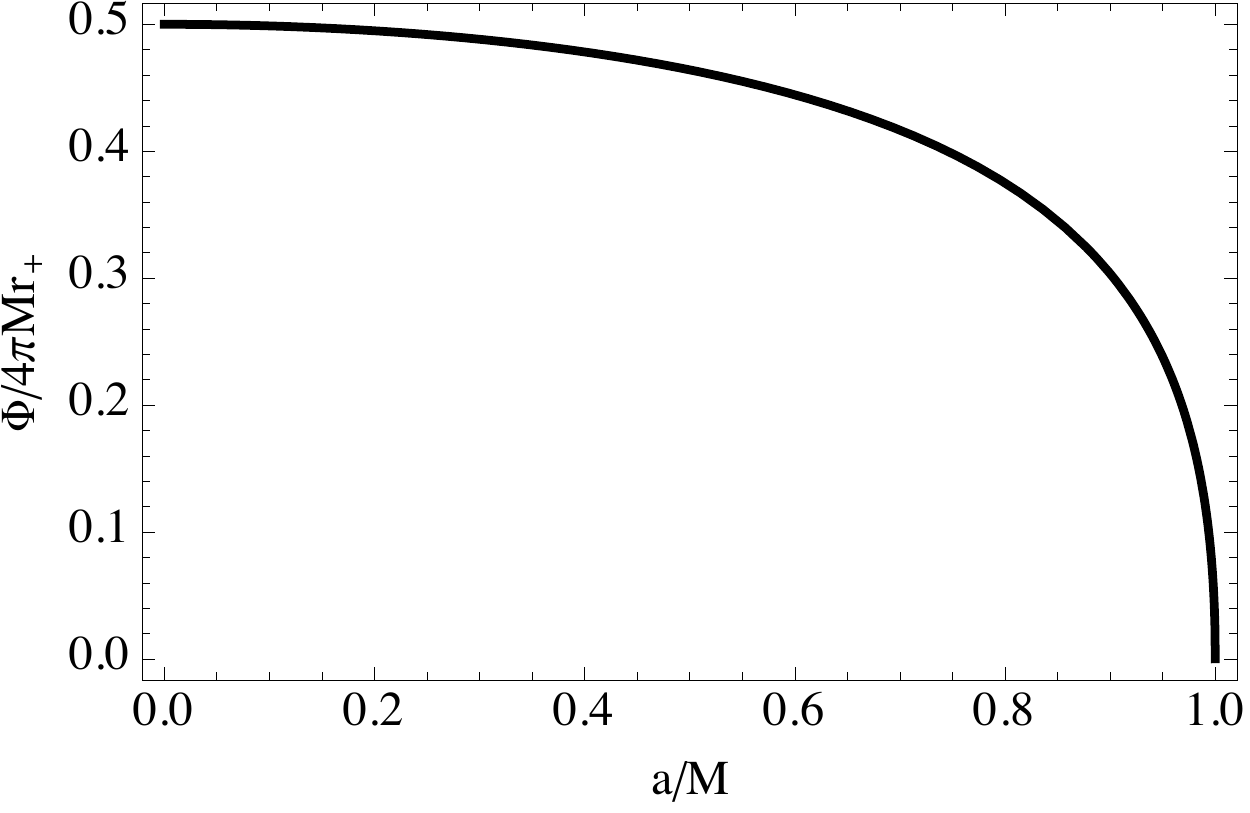}
\caption{The flux threading the northern hemisphere of the black hole horizon drops continuously as $a/M\rightarrow 1$.  We have set $B=1$.}
\label{fig:waldfluxes}
\end{center}
\end{figure*}

If the field was only expelled at the extremal limit, one could dismiss the effect as a pathology of $a/M=1$.  It is impossible to achieve extremal spins in nature, so the implications of the Meissner effect would be limited.  However,  the  flux is expelled in a continuous way as the black hole is spun up.  It is not a discontinuous effect that only appears exactly at $a/M=1$.  The flux threading the northern hemisphere of the horizon is
\beq
\Phi = 2 \pi \int_0^{\pi/2} F_{\theta\phi} d\theta.
\eeq
The integral is restricted to one hemisphere of the horizon because the flux over the entire horizon is trivially zero (because the magnetic monopole charge of the black hole is zero).  Plugging in the Wald solution gives \cite{1974PhRvD..10.1680W,1975PhRvD..12.3037K}
\beq
\Phi = \pi r_+^2 B (1-a^4/r_+^4),
\eeq
where $r_+=M+\sqrt{M^2-a^2}$ is the radius of the horizon and $B$ is the field strength at infinity.  Figure \ref{fig:waldfluxes} shows how $\Phi$ drops as the black hole is spun up.   The drop is partly coming from the fact that the black hole is shrinking as it spins up.  This contribution is not particularly interesting, as even in flat space the flux of a uniform field through a sphere depends on its surface area.  However, the area-normalized flux, $\Phi/(4 \pi M r_+)$, also drops with spin (the area of the northern hemisphere of the horizon is $4 \pi M r_+$). 

One might worry that the Meissner effect relies on a peculiar feature of the Wald solution. This solution is a test field, but the effect persists for non-test fields \cite{bicak1989,1991JMP....32..714K,1998PhRvD..58h4009C,2000PhyS...61..253K}.  The Wald solution is a vacuum field and the vector potential is a Killing vector, so it is a very special configuration.  
However, Bi{\v c}{\'a}k and Dvo{\v r}{\'a}k \cite{1976GReGr...7..959B} have found solutions which vastly generalize the Wald solution and their solutions also display the Meissner effect \cite{1985MNRAS.212..899B}.  They found a general multipole expansion which can be adapted to (almost) any axisymmetric, stationary magnetic field in the Kerr metric, including non-vacuum fields sourced by current distributions.  
Their result is often summarized as proving \cite{2007IAUS..238..139B}
\begin{proposition*}
``All stationary, axisymmetric magnetic fields are expelled from the Kerr horizon as $a/M\rightarrow 1$.''
\end{proposition*}
The standard Blandford-Znajek (BZ) model \cite{bz77} of spin-powered black hole jets is stationary and axisymmetric, so this result appears to rule out BZ jets at high spins.

Our first observation is that stationary and axisymmetric fields which become entirely radial near the horizon are not expelled by the Meissner effect and may continue to power jets up to the extremal limit. This is an important possibility because early work on the BZ model and recent simulations both suggest
the fields of black hole jets have a large split monopole component
\cite{bz77,1979AIPC...56..399L,1982MNRAS.198..345M,phinney1983,2012MNRAS.423.3083M,2012JPhCS.372a2040T,2013MNRAS.436.3741P}.  This provides a natural mechanism to power jets from Cyg X-1 and GRS 1915+105.  It appears to be consistent with the perturbative solutions constructed by \cite{2011MNRAS.412.2417T}, which describe slowing rotating fields threading extremal Reissner-Nordstr{\" o}m horizons.

One can imagine embedding the Bi{\v c}{\'a}k and Dvo{\v r}{\'a}k solutions in a conductive magnetosphere.  Conductivity does not enter into Maxwell's equations, so for a fixed current distribution, turning on conductivity does not affect whether the field lines thread the horizon.  However, a conductive magnetosphere may redistribute the current.  The final current distribution might thread the horizon with flux even if the initial current distribution did not (or vice versa).  Simulations suggest conductive magnetospheres choose current distributions which evade the Meissner effect \cite{2007MNRAS.377L..49K}.  The result of \cite{1985MNRAS.212..899B} suggests such current distributions must be nonstationary or nonaxisymmetric.  Our observation is that the field may remain stationary and axisymmetric (as in the original BZ model) provided it becomes radial at the horizon.

One might argue that a sufficiently powerful accretion disk can drag any field onto an extremal horizon despite the Meissner effect.  We give a simple geometrical reason why this is impossible unless the field becomes radial at the horizon.  It has been argued that jets can be powered directly by the ergosphere, so that even if the Meissner effect operated it would not be relevant for BZ jets \cite{1990ApJ...354..583P,2007MNRAS.377L..49K}.  We argue that this suggestion is incorrect unless the BZ model is significantly modified.

Finally we note that in our simulations of black hole jets there is no evidence for the Meissner effect.  We have observed previously that the fields in our simulations have a large split monopole component \cite{2013MNRAS.436.3741P}.  So this is consistent with our observation that  fields which become radial at the horizon evade the Meissner effect.  Furthermore, simulations generate split monopole fields spontaneously, which suggests this mechanism is naturally occurring.

Our paper is organized as follows.  In Sec. \ref{sec:jets} we review the physics underlying the black hole Meissner effect and debunk two proposals for evading the Meissner effect.  In Sec \ref{sec:evade} we discuss three ways the black hole Meissner effect can be evaded.  Of these, split-monopole fields provide a particularly natural solution.  In Sec. \ref{sec:conc} we summarize and conclude.

\section{The Meissner Effect and Jets}
\label{sec:jets}

\subsection{Black Hole Jets}

Black hole jets may be powered by the black hole's spin or by an accretion flow.  The standard model of spin-powered jets is the Blandford-Znajek (BZ) model \cite{bz77,1986bhmp.book.....T}.     It describes how magnetic field lines thread the horizon and extract the black hole's rotational energy.  The jet power is 
\beq\label{eq:pbz}
P_{\rm jet} = \frac{1}{8\pi} \Omega_H^2 \Phi^2,
\eeq
where $\Omega_H=a/(2Mr_+)$ is the angular velocity of the horizon.  Recent observations of jets from galactic X-ray binaries are consistent with \eqref{eq:pbz} \cite{2012MNRAS.419L..69N,2013ApJ...762..104S,2013SSRv..tmp...73M}.  Numerous general relativistic magnetohydrodynamic (GRMHD) simulations  have checked and verified the BZ model \cite{
2001MNRAS.326L..41K,
2004MNRAS.350.1431K,
2005MNRAS.359..801K,
2006MNRAS.368.1561M,
2007MNRAS.377L..49K,
2009ApJ...699.1789T,
2009MNRAS.394L.126M,
2010ApJ...711...50T,
2011MNRAS.418L..79T,
2012MNRAS.423.3083M,
2012JPhCS.372a2040T,
2013MNRAS.436.3741P}.

It is usually assumed that the flux threading the horizon, $\Phi$, is fixed by the accretion rate (e.g. \cite{2003PASJ...55L..69N}).  The field builds up until its outward pressure balances the inward ram pressure of the accretion flow.  Numerous GRMHD simulations support this picture.  So  jet power is expected to scale with spin as $P_{\rm jet}\sim \Omega_H^2$, for fixed accretion rate.

This simple picture could break down for rapidly rotating black holes if the black hole Meissner effect expels the magnetic field  and prevents it from reaching its equilibrium value \cite{1985MNRAS.212..899B,1991NYASA.631..235P,2007IAUS..238..139B}.  In this case, jet power would drop off at high spins and go to zero at the extremal limit.  

\subsection{What Causes the Meissner Effect?}
\label{sec:cylinder}

One might think that a sufficiently powerful accretion disk can drag any field geometry onto a black hole horizon, even in the extremal limit.  It turns out this is impossible.  To explain why, we turn to the physics underlying the black hole Meissner effect. 

%All black holes have $g_{rr}=\infty$ at $r=r_+$. (The radial component of the metric blows up at the horizon in Boyer-Lindquist coordinates).  However, the horizons of extremal black holes are especially badly behaved.  
In the extremal limit,
\beq\label{eq:rprop}
\int_{r_+}^{r_+ + \epsilon} \sqrt{g_{rr}}dr \rightarrow \infty,
\eeq
and the proper length of the black hole throat blows up.  
Note that this is a special feature of extremal black holes.  All black holes have infinite $g_{rr}$ at the horizon, but the integral \eqref{eq:rprop} is infinite only for extremal black holes.
  
\begin{figure}[!ht]
\begin{center}
\includegraphics[scale=1]{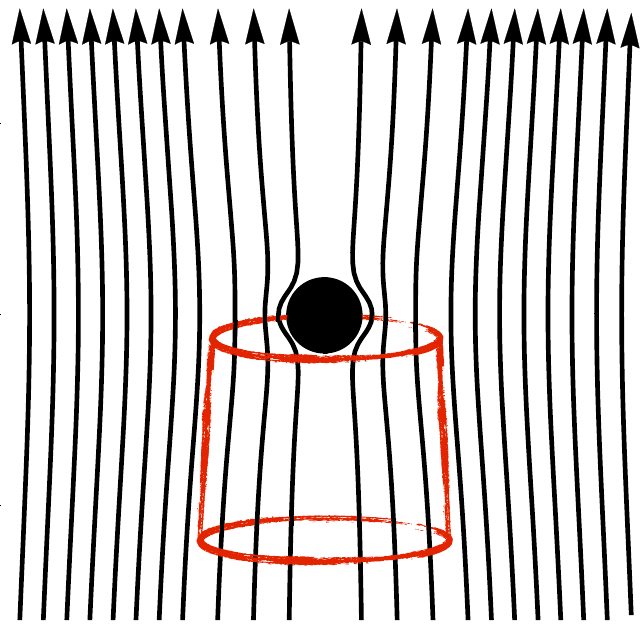}
\caption{The $a/M = 1$ Wald solution of Figure \ref{fig:waldfields}.}
\label{fig:cylinder}
\end{center}
\end{figure}

Now consider the red cylinder in Figure \ref{fig:cylinder}.  In the extremal limit, the surface area of the top of the cylinder blows up but the surface area of the bottom stays finite.  So to maintain $\nabla\cdot \mathbf{B}=0$, the magnetic field must vanish at the horizon in the extremal limit.  This causes the Meissner effect.  It makes no difference whether or not there is an accretion disk.  If a disk (or anything else) were to forcibly drag the field onto the horizon, then the flux through the top of the cylinder would be infinite and  $\nabla \cdot \mathbf{B}=0$ would be broken.  Invoking an accretion disk does not help.   

For concreteness, we give the cylinder argument in the zero angular momentum observer (ZAMO) frame.  The basis vectors are \cite{1972ApJ...178..347B}
\begin{align}
\mathbf{e}_{\hat{t}} &= \left(\frac{A}{\rho^2 \Delta}\right)^{1/2}\frac{\partial}{\partial t} 
	+ \frac{2Mar}{\left(A\rho^2\Delta\right)^{1/2}}\frac{\partial}{\partial \phi}, \\
\mathbf{e}_{\hat{r}} &= \frac{\Delta^{1/2}}{\rho}\frac{\partial}{\partial r},\\
\mathbf{e}_{\hat{\theta}} &= \frac{1}{\rho}\frac{\partial}{\partial \theta},\\
\mathbf{e}_{\hat{\phi}} &= \frac{\rho}{A^{1/2}\sin\theta}\frac{\partial}{\partial \phi},
\end{align}
where
\begin{align}
\Delta &= r^2-2Mr + a^2, \\
\rho^2 &= r^2 + a^2\cos^2\theta,\\ 
A &= (r^2+a^2)^2 -a^2 \Delta \sin^2\theta.
\end{align}
The one-forms are
\begin{align}
\mathbf{e}^{\hat{t}} &= \left(\frac{\rho^2 \Delta}{A}\right)^{1/2}dt, \\
\mathbf{e}^{\hat{r}} &= \frac{\rho}{\Delta^{1/2}}dr,\\
\mathbf{e}^{\hat{\theta}} &= \rho d\theta,\\
\mathbf{e}^{\hat{\phi}} &= -\frac{2Mar \sin\theta}{\rho A^{1/2}}dt
+\frac{A^{1/2}}{\rho}\sin\theta d\phi.
\end{align}
The magnetic field is
\beq
B^{\hat{i}} = *F^{\hat{i}\hat{t}}, \quad (\hat{i}=\hat{r},\hat{\theta},\hat{\phi}).
\eeq
$B^{\hat{i}}$ is a three-vector living on a $t={\rm const.}$ slice of the Kerr metric.  The no-monopoles constraint is \cite{1986bhmp.book.....T}
\beq\label{eq:nomono}
\nabla \cdot \mathbf{B} = 0.
\eeq
The components of the connection are 
\beq
\Gamma_{\hat{i}\hat{j}\hat{k}}
=-\gamma^{\hat{l}}_{\hat{j}\hat{k}} g_{\hat{i}\hat{l}}
-\gamma^{\hat{l}}_{\hat{k}\hat{i}} g_{\hat{j}\hat{l}}
+\gamma^{\hat{l}}_{\hat{i}\hat{j}} g_{\hat{k}\hat{l}},
\eeq
where the commutator coefficients are defined by
\beq
[\mathbf{e}_{\hat{j}},\mathbf{e}_{\hat{k}}] = \gamma^{\hat{i}}_{\hat{j}\hat{k}} \mathbf{e}_{\hat{i}}.
\eeq

Integrating \eqref{eq:nomono} over the cylinder gives
\beq
\int_{C} \nabla \cdot \mathbf{B} \thinspace \mathbf{e}^{\hat{r}}\wedge \mathbf{e}^{\hat{\theta}}\wedge  \mathbf{e}^{\hat{\phi}}  = 0 .
\eeq
This is the same as the flux through the surface of the cylinder, by the divergence theorem.  The contribution from the top of the cylinder is
\beq
\int_{\partial C} B^{\hat{\theta}} \thinspace \mathbf{e}^{\hat{r}}\wedge  \mathbf{e}^{\hat{\phi}}
=  \int_{\partial C} B^{\hat{\theta}}  \left(\frac{A}{\Delta}\right)^{1/2} \sin\theta drd\phi.
\eeq
Assume $B^{\hat{\theta}}$ is axisymmetric and stationary.  Then at the extremal limit, the right hand side is finite only if $B^{\hat{\theta}}$ goes to zero at the horizon faster than $\sqrt{\Delta(r)}$.   This implies the Meissner effect for most stationary, axisymmetric fields (regardless of whether or not there is an accretion disk).   The one exception are fields which lie entirely along $\mathbf{e}^{\hat{r}}$.  We will revisit this exception later.   

The Meissner effect could also be evaded by a field which threads the horizon but then drops to zero as it rises up out of the throat (we discuss such an example in Section \ref{sec:other}).  However, such a field could not power conventional jets, because jets require field lines which extend from the horizon to a distant load region.  

We have depicted the cylinder argument at the equatorial plane, but it immediately generalizes to all polar angles because the length of the black hole throat \eqref{eq:rprop} blows up at all polar angles.

The cylinder argument is appropriate for observers who remain outside the black hole.  From the perspective of infalling observers, the length of the black hole throat remains finite in the extremal limit.  It is not clear why the Meissner effect should exist at all from this perspective, although clearly it does.  

We have shown that the field goes to zero at Boyer-Linquist radius $r/M=1$.  In these coordinates, the region between the horizon and the innermost stable circular orbit are all at $r/M=1$ in the extremal limit. A closer look at the near horizon geometry \cite{1972ApJ...178..347B} shows that the field must go to zero at the photon orbit, an infinite (spacelike) distance outside the horizon.  So the Meissner effect is somewhat stronger than suggested above.

\subsection{The Importance of the Horizon}

%It has been claimed that jets would not be quenched even if the Meissner effect prevented the field from threading the horizon \cite{2007MNRAS.377L..49K} (despite the fact that fields do thread the horizon in all detailed models of spin-powered jets constructed to date).   We disagree with this claim.  
%This claim is based on the belief that the horizon plays no essential role in the energy extraction process.  This belief is based on the following argument.  There are spacetimes which have no horizons and yet have ergospheres and the Penrose energy extraction process.  Examples include the naked singularities describing over-spinning Kerr black holes and the spacetimes considered by \cite{2012MNRAS.423.1300R}.  These spacetimes have no horizon, so clearly the horizon plays no essential role in the Penrose process in these spacetimes.  This suggests the horizon plays no essential role in this process in any spacetime.
%We will now argue that this is incorrect.  

In black hole spacetimes, the horizon is crucially connected to the energy extraction process, so that if the Meissner effect operated, ordinary BZ jets would be impossible.  To motivate this claim, consider the original Penrose process \cite{1971NPhS..229..177P} in horizon penetrating coordinates.  A negative energy particle is created in the ergosphere. All negative energy geodesics eventually cross the horizon \cite{1984GReGr..16...43C}.  Suppose a ``wall'' mimicking the Meissner effect were to prevent the particle from crossing the horizon indefinitely.  Then the particle would need to be boosted back to positive energy.  So no net energy would be extracted in the end.  Roughly speaking, the work done by the wall negated the energy extraction.  

A similar claim holds for fields.  If the Meissner effect or some other wall keeps the fields out of the horizon, then there is no net energy extraction.   In the Meissner effect case, the wall is present the whole time.  It would be acausal for energy to be extracted for awhile and then at some point returned to the black hole.  What happens is that energy is not extracted at any point.  The way to think of it is that a field configuration which is not on a horizon crossing trajectory cannot be a negative energy field configuration (just as a particle which is not on a horizon crossing geodesic cannot be a negative energy particle).  If the Meissner effect operates, then there are no horizon crossing field configurations, so there is no energy extraction and ordinary BZ jets are impossible.

This is a consistency condition in black hole spacetimes which is not present in spacetimes without horizons.  The consistency condition is that field lines must cross the horizon (possibly at some point in the future) for jets to extract the black hole's rotational energy.  This can be understood intuitively and proved rigorously using the membrane paradigm.  The membrane paradigm replaces the black hole interior with a membrane living on the boundary of the interior (the horizon).  The black hole's rotational energy is stored on the membrane.  Jets are powered because field lines torque the membrane and extract its rotational energy (as in the BZ model).  So if field lines cannot thread the horizon,  there can be no torque on the horizon and there are no BZ jets.  

Jets are possible even if the field threads the horizon in the distant future, after the energy has been extracted.  This is because the membrane's response is dictated by a teleological Green's function, reflecting the global nature of event horizons.  The membrane's ``response'' precedes the torque.  In extreme cases (e.g., in Boyer-Lindquist coordinates), the external field need not reach the horizon until $t\rightarrow \infty$.  However, if the Meissner effect operates, then field lines are prevented from threading the horizon even as $t\rightarrow \infty$, so there is no torque on the horizon and there cannot be BZ jets.

We have argued that negative-energy field configurations must be horizon-crossing.  The magnetic field could be part of a larger negative energy configuration, in which case the magnetic field itself need not cross the horizon.  Some part of the larger configuration would need to cross the horizon for there to be net energy extraction.  For example, the magnetic field could be coupled to a fluid.  Energy extraction would be possible even if only the fluid crossed the horizon.  In the membrane formulation of this process, the jet would be powered by hydrodynamic (rather than magnetic) torques acting on the horizon.  In our simulations, jets are powered by magnetic torques as envisioned in the original BZ model \cite{2013MNRAS.436.3741P}. 

The consistency condition we have been discussing and the membrane paradigm cannot be formulated in spacetimes without horizons, such as naked singularities and the spacetimes considered by \cite{2012MNRAS.423.1300R}. 

Our argument relies on the membrane paradigm as a complete and exact description of black holes with horizons.  We have shown that it correctly describes black hole jet simulations \cite{2013MNRAS.436.3741P}.   More generally, the action principle formulation of Parikh and Wilczek \cite{1998PhRvD..58f4011P} makes it clear that the membrane paradigm is a complete and exact description of all black hole physics (at least for classical observers outside the black hole).  Let us expand on this point.    Electrodynamics is described by an action,
\beq\label{eq:action}
S= \int d^4 x \mathcal{L}.
\eeq  
On a black hole spacetime, we may split the action into
\beq\label{eq:stot}
S = S_{\rm in} + S_{\rm out},
\eeq
where $S_{\rm in}$ ($S_{\rm out}$) is the action obtained by restricting the integral in \eqref{eq:action} to the black hole interior (exterior).  An observer outside the black hole has no access to $S_{\rm in}$, but varying $S_{\rm out}$ alone does not give the right physics because it is not stationary on solutions of  $\delta S =0$.  So we define the external observer's effective action
\beq\label{eq:seff}
S_{\rm eff} = S_{\rm out} + S_{\rm mb},
\eeq
by supplementing $S_{\rm out}$ with a correction term, $S_{\rm mb}$, defined such that $\delta S_{\rm eff}=0$ and $\delta S=0$ agree:
\beq\label{eq:duality}
\delta S_{\rm eff} = \delta S.
\eeq
For electrodynamics, the required correction term is  \cite{1998PhRvD..58f4011P} 
\beq\label{eq:smb}
S_{\rm mb} = \int d^3x \sqrt{-h} j_s \cdot A,
\eeq
where 
\beq
j_s^i = F^{ij}n_k
\eeq
is the membrane current, $A$ is the vector potential, $F=dA$, and $n^i$ is the outward-pointing, space-like unit normal to the horizon.  The surface integral in \eqref{eq:smb} is over the horizon.  The horizon now carries a current.  Enforcing regularity at the horizon leads to Ohm's law and the horizon resistance 377 ohms (see \cite{1998PhRvD..58f4011P}).

Comparing \eqref{eq:stot} and \eqref{eq:seff}, we see that an action over the black hole interior, $S_{\rm in}$, has been traded for an action over the black hole horizon, $S_{\rm mb}$.    In this sense, the membrane paradigm gives a holographically dual description of the black hole interior.  The original action and the effective action have the same variation, so they are physically equivalent (at least for classical observers outside the black hole).  For jet model-building, the membrane paradigm has the advantage that the black hole's rotational energy is stored in a well-defined place: it is located on the horizon (see \cite{1986bhmp.book.....T}).  This makes it clear that magnetic fields must thread the horizon to power jets in the membrane paradigm, and so they must thread the horizon in any formulation of jet physics, by the duality \eqref{eq:duality}.  It is not sufficient for field lines to thread the ergosphere alone in black hole spacetimes.

For completeness, we explain in more detail how the membrane action \eqref{eq:smb} was derived by \cite{1998PhRvD..58f4011P}.   The electromagnetic Lagrangian is $\mathcal{L}=\mathcal{L}(A_\mu,\partial_\mu A_v)$. Variation of the exterior piece of the action alone gives
\begin{align}
\delta S_{\rm out} &=
\int d^4x \sqrt{-g} \thinspace \delta A_\mu 
\left(\frac{\partial \mathcal{L}}{\partial A_\mu}
	-\partial_\nu \frac{\partial \mathcal{L}}{\partial(\partial_\nu A_\mu)}\right)\notag\\
	&+\int d^3x \sqrt{-h} \thinspace n_\mu \left(\frac{\partial\mathcal{L}}{\partial(\partial_\mu A_\nu)}\delta A_\nu\right),
\end{align}
where we have used integration by parts and the divergence theorem.  The first integral on the RHS gives the usual Maxwell's equations in the black hole exterior.  It is stationary on solutions of the total action, $\delta S=0$.  The second integral on the RHS  is a surface term supported on the black hole horizon.  It  does not vanish on solutions of $\delta S=0$ and must be canceled.  The correction term we need is
\beq\label{eq:scorr}
\delta S_{\rm mb} = -\int d^3x \sqrt{-h} \thinspace n_\mu \left(\frac{\partial\mathcal{L}}{\partial(\partial_\mu A_\nu)}\delta A_\nu \right).
\eeq
Plugging in the electromagnetic Lagrangian, $\mathcal{L} = -F^2/4 + J\cdot A$, 
gives the membrane action \eqref{eq:smb}, as claimed.  All of the black hole membrane paradigm flows from this starting point.  Note that this formulation of the membrane paradigm is entirely covariant; it can be adapted to any coordinate system or frame, so long as the observer remains outside the black hole.

\subsection{Four Interpretations of Black Hole Jets}

\begin{figure}
\includegraphics[width=\columnwidth]{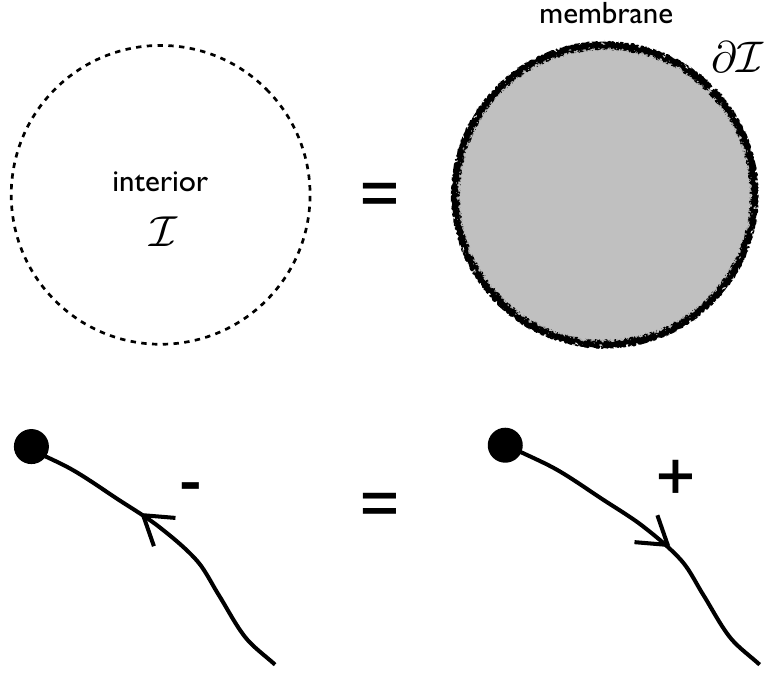}
\caption{\emph{Top panel:} The membrane paradigm replaces the black hole interior with a membrane living on the boundary of the interior (the horizon).  Both pictures give a complete and exact description of black hole physics (at least for classical observers outside the black hole).  \emph{Bottom panel:} Black hole energy extraction may be described either as negative energy entering the black hole or as positive energy leaving the black hole.  So there are four equivalent ways to describe how jets are powered: positive energy flows out of the interior, negative energy flows into the membrane, etc.}
\label{fig:holo}
\end{figure}

One is free to say that energy is exchanged with the black hole interior or with a membrane living on the black hole horizon.   One may further choose whether to describe the process as negative energy flowing into the black hole (e.g., Boyer-Lindquist coordinates), or as positive energy flowing out of the black hole (e.g., in a local frame).  So there are four physically equivalent interpretations of spin-powered jets. Figure \ref{fig:holo} summarizes the possibilities.

Note that all negative energies in this sense are ``energies at
infinity.''  No local observer can measure a negative energy (by the
energy theorems for ordinary matter).  Negative energies can at best
be inferred indirectly, for example, by scattering a particle through
the Penrose process and then inferring that the intermediate,
infalling particle had negative energy based on the change in energy
of the scattered particle.

When black hole spin energy is extracted, an observer at infinity may
infer that negative energy particles are flowing into the black hole.
A local observer will find only positive energy particles.  In the
local observer's reference frame, the particles with negative energies
at infinity will appear to be positive energy particles flowing out of
the black hole.  But of course all of the particles and the observer
are falling into the black hole, it is only their relative motion that
causes it to appear to the local observer that some particles are
flowing out.

\subsection{GRMHD Simulations}

One might try to understand the Meissner effect using GRMHD simulations.  It is not clear whether this is a reliable approach.  Simulations discretize space into grid cells.  Grid cells at the horizon have small coordinate sizes, typically $\Delta r/M \approx 0.005$, but they have infinite proper lengths \eqref{eq:rprop} in the extremal limit. 

\begin{figure}
\includegraphics[width=\columnwidth]{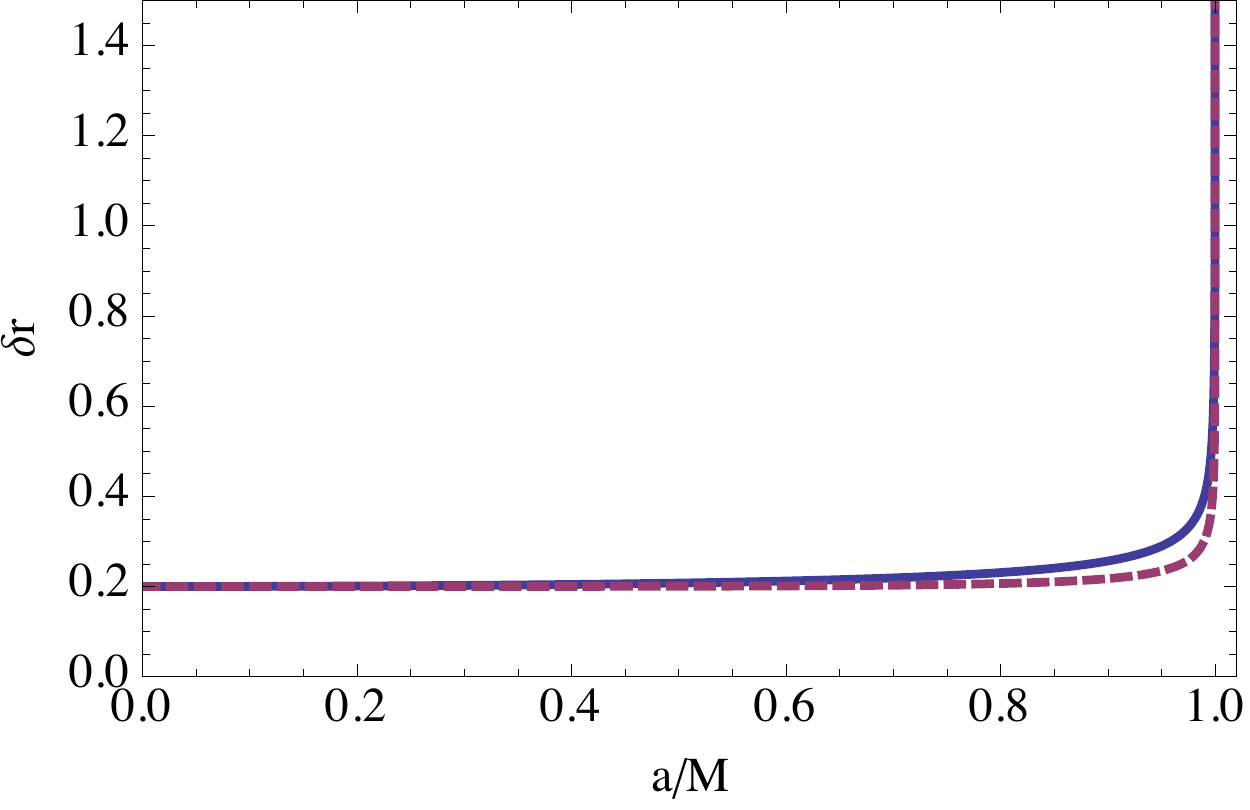}
\caption{Proper lengths, $\delta r = \int_{r_+}^{r_++\Delta r} \sqrt{g_{rr}}dr$, of grid cells at $r=r_+$ as a function of black hole spin. Curves correspond to grid cells at polar angles  $\theta=0$ (solid blue) and $\theta=\pi/2$ (dashed red). We assume a typical GRMHD simulation, for which the coordinate length of these grid cells is $\Delta r = 0.005$ (for example, the simulations of \cite{2013MNRAS.436.3741P}).   For $a/M \gtrsim 0.95$, the proper lengths blow up and the black hole throat is not resolved by the simulation.}
\label{fig:fuzzywald}
\end{figure}

The simulation's horizon is outside the true horizon.  The separation is of order the grid cell size.  So  simulated black holes do not have infinitely long throats in the extremal limit.  Most of the throat is inside a single grid cell at the horizon.  The region of spacetime that causes the Meissner effect is unresolved (see section \ref{sec:cylinder}).  This becomes a serious limitation for $a/M \gtrsim 0.95$ (see Figure \ref{fig:fuzzywald}).     So it is not clear whether GRMHD simulations can give reliable results on the Meissner effect for rapidly rotating black holes. (The surface area of the top of the red cylinder in Figure \ref{fig:cylinder} never blows up.)  Of course most other aspects of black hole jets and accretion physics are insensitive to the proper length of the throat and can be simulated reliably.  The infinite length of the throat is in a spacelike direction while accreting gas follows timelike curves and reaches the horizon in finite proper time.  However, magnetic field lines are spacelike curves and experience the proper length of the throat through the Meissner effect.
 
It is worth noting that for spins $a/M<0.95$, our GRMHD simulations show no sign of the Meissner effect.  
In an earlier paper \cite{2013MNRAS.436.3741P}, we used the GRMHD code HARM \citep{2003ApJ...589..444G,2006MNRAS.367.1797M} to evolve a three dimensional, magnetized, turbulent accretion disk in the Kerr metric.  We work in horizon penetrating coordinates, with a logarithmically spaced radial grid that concentrates resolution on the inner regions of the accretion flow.  As the disk drags magnetic fields onto the black hole, jets develop spontaneously.  The jets are correctly described by the BZ model in its membrane paradigm formulation \cite{2013MNRAS.436.3741P}.  A full description of the setup can be found in \cite{2013MNRAS.436.3741P}.

\begin{figure}
\includegraphics[width=\columnwidth]{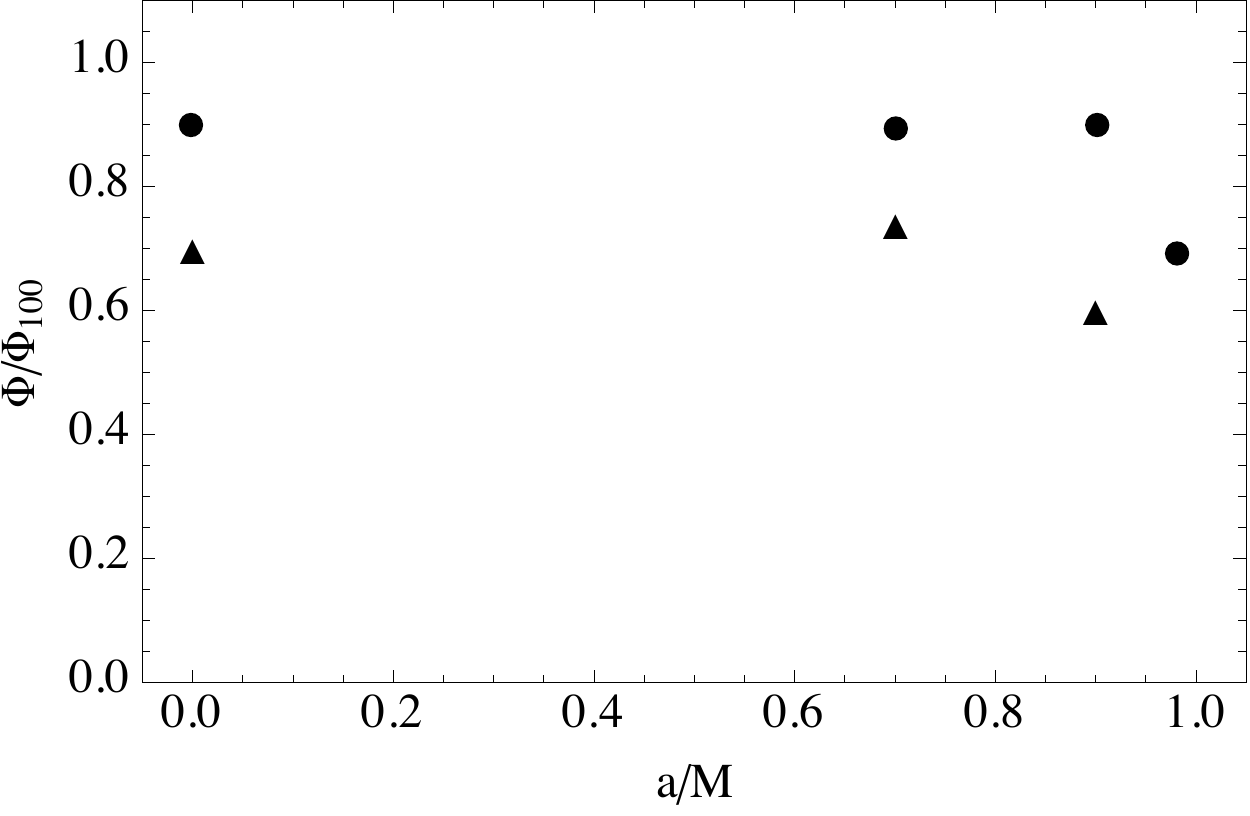}
\caption{Flux threading the northern hemisphere of the horizon as a function of spin for seven of our GRMHD simulations \cite{2013MNRAS.436.3741P}.  Symbols indicate whether the magnetic field in the GRMHD initial conditions is a single poloidal loop (triangles) or a series of poloidal loops (circles).  The flux is time averaged over the steady-state period of the simulations and normalized to the flux threading the northern hemisphere of a sphere at $r/M=100$.}
\label{fig:fluxvsspin}
\end{figure}

Figure \ref{fig:fluxvsspin} shows the flux threading the northern hemisphere of the black hole horizon, $\Phi$, as a function of black hole spin for a series of these simulations.  The data has been time averaged over the steady-state period of the simulations and  normalized to the flux threading the northern hemisphere of a sphere at $r/M=100$. The horizon's equilibrium flux  shows no significant dependence on spin.  It is set entirely by the accretion flow, as expected.  Of course the Meissner effect is only expected to be important for $a/M>0.95$, which is  the regime where simulations do not resolve the black hole throat.

\section{Evading the Meissner Effect}
\label{sec:evade}

In this section we describe three ways to evade the Meissner effect.  The first provides a particularly natural mechanism for powering astrophysical jets.

\subsection{Split Monopole Fields}

The cylinder argument of section \ref{sec:cylinder} rules out the possibility of stationary axisymmetric fields threading the horizon in the extremal limit unless the field becomes entirely radial at the horizon.  Fields entirely along $dr$ run parallel to the throat. It is impossible to adapt the cylinder argument to these solutions. 

The simplest example of such a field is a magnetic monopole.  Of course, there are no magnetic monopoles in astrophysics.  However, split monopoles occur in black hole jets.  A split monopole is the same as a monopole except the sign of the field is flipped on one side of the equatorial plane.  The field is supported by currents flowing in the plane.  The currents are supported by an accretion disk. 
Early work on the BZ model and recent GRMHD simulations suggest black hole jets have a large split monopole component \cite{bz77,1979AIPC...56..399L,1982MNRAS.198..345M,phinney1983,2012MNRAS.423.3083M,2012JPhCS.372a2040T,2013MNRAS.436.3741P}.   For example, in our GRMHD simulations \cite{2013MNRAS.436.3741P}, a turbulent accretion disk brings magnetic fields to the black hole, jets develop spontaneously, and the fields powering the jets have a large split monopole component.

\begin{figure}[!ht]
\begin{center}
\includegraphics[width=\columnwidth]{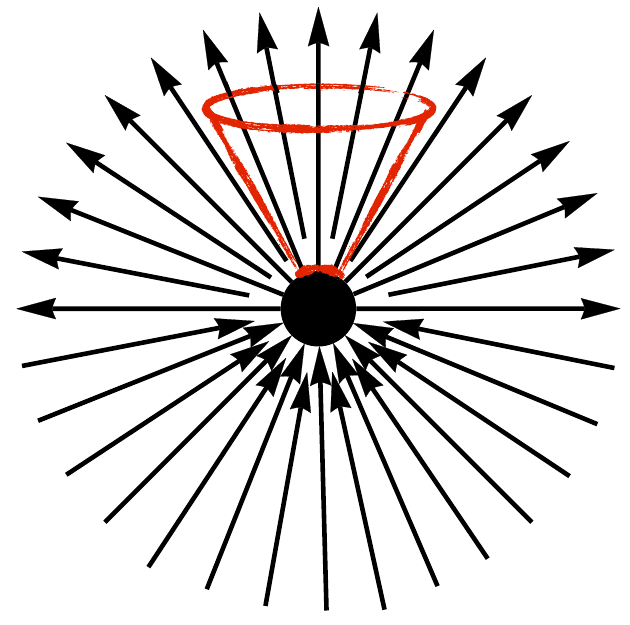}
\caption{A black hole with a split monopole magnetic field.  The field is supported by currents flowing in the equatorial plane.  The areas of the top and bottom faces of the red cylinder remain finite in the extremal limit.  So a split monopole can thread the horizon without breaking $\nabla\cdot\mathbf{B}=0$, even as the black hole throat becomes infinitely long.  In other words, there is no Meissner effect for split monopole fields.  They may continue to power jets at the extremal limit.}
\label{fig:splitmono}
\end{center}
\end{figure}

Figure \ref{fig:splitmono} shows a black hole with a split monopole field.  In this picture the top and bottom faces of the red cylinder are orthogonal to $dr$.  The surface area remains finite in the extremal limit, so the field can thread the horizon while maintaining $\nabla \cdot \mathbf{B}=0$.  There is no tendency to expel the field.  

This appears to contradict the result of Bi{\v c}{\'a}k and Janis \cite{1985MNRAS.212..899B}, which is often understood as ruling out the possibility of axisymmetric stationary fields threading the extremal horizon.  However, a closer look at their result shows that there is no monopole component in their multipole expansion of the field.  They did not investigate current distributions which extend down to the event horizon, which precludes the possibility of a split monopole.  So there is no contradiction.  %Identifying the monopole loophole gives a theorem:
%dipole and higher order moments of stationary, axisymmetric magnetic fields are expelled from the Kerr horizon as $a/M\rightarrow 1$.

It is perhaps surprising that dipole fields are expelled while split monopole fields are not.  At the horizon they are very similar, each is positive over one hemisphere and negative over the other.  Split monopole vector potentials have $A_\phi \sim \cos\theta$ while dipole potentials have $A_\phi \sim \sin^2\theta$.  If the split monopole could be expanded as a sum of dipole and higher order multipole moments then it would be expelled.  The reason this is impossible is that the split monopole $A^\phi ~\sim \cos\theta/\sin^2\theta$ has Dirac string singularities at $\theta = 0,\pi$.  Only smooth vector potentials can be expanded as the sum of dipole and higher order multipoles. 

More explicitly, consider the three dimensional Maxwell's equations on the horizon,
\beq\label{eq:hodge}
\Delta_H A^i \equiv {F^{ij}}_{;j} = J^i,
\eeq
where $A^i$ and $F^{ij}$ are the horizon gauge fields as defined in the black hole membrane paradigm.  Assume a stationary vector potential and set $a/M=0$, for simplicity.  Then equation \eqref{eq:hodge} is a differential equation on the two-sphere and $\Delta_H$ is the Hodge Laplacian on $S^2$.  The multipole moments of the field are eigenvectors of $\Delta_H$.

The eigenfunctions of $\Delta_H$ are just the usual spherical harmonics,
\beq
r_+^2 \Delta_H Y^m_\ell(\theta,\phi) = -\ell(\ell+1)Y^m_\ell(\theta,\phi),
\eeq
where $\ell \geq 0$ and $-\ell \leq m \leq \ell$.

The eigenforms of $\Delta_H$ are $dY^m_\ell$ and $*dY^m_\ell$, because
\begin{align}
r_+^2 \Delta_H(dY^m_\ell) &= r_+^2 (dd^* + d^*d)(dY^m_\ell) \notag\\
					&= -\ell(\ell+1) dY^m_\ell, \notag\\
r_+^2 \Delta_H(*dY^m_\ell) &= r_+^2 (dd^* + d^*d)(*dY^m_\ell)  \notag\\
						&= -\ell(\ell+1) *dY^m_\ell,\notag
\end{align}
where $d^*= - *d*$.    (Since the sphere has vanishing first cohomology group, $H^1(S^2)=0$, this is a complete set of eigenforms.)  The $\ell=0$ mode is the zero form, so we may restrict attention to $\ell \geq 1$.  In other words, the dipole and higher order moments give a complete set of eigenvectors.  Any smooth vector potential may be expanded in this basis.  All such fields are expelled in the extremal limit.  The split monopole potential is singular, so it may not be expanded in this basis.  This is why it evades the Meissner effect.

%The Hodge and connection Laplacians, $\Delta_H$ and $\Delta$, are related via \eqref{eq:current2}:
%\beq
%r_+^2 \Delta _H A_i  = A_i - r_+^2 \Delta A_i.
%\eeq
%So the eigenvectors of the Hodge Laplacian with eigenvalue $-\ell(\ell+1)/r_+^2$ are eigenvectors of the connection Laplacian with eigenvalue $(1+\ell+\ell^2)/r_+^2$.

The axisymmetric eigenvectors are particularly simple, so we record them here.  They have components
\begin{align}
(*dY^0_\ell)_\theta  &= 0,\\
(*dY^0_\ell)_\phi  &= \sqrt{g} g^{\theta\theta} \frac{\partial Y^0_\ell}{\partial \theta}.
\end{align}
An orthonormal set is 
\beq
A_{\ell\phi} = \sin\theta Y^1_\ell(\theta,0),
\eeq
for $\ell\geq 1$.  The first few are
\begin{align}
A_{1\phi} &= \frac{1}{2}\sqrt{\frac{3}{2\pi}} \sin^2 \theta,\\
A_{2\phi} &= \frac{1}{2}\sqrt{\frac{15}{2\pi}} \cos\theta \sin^2 \theta,\\
A_{3\phi} &= \frac{1}{8}\sqrt{\frac{21}{\pi}} (5\cos^2\theta -1)\sin^2 \theta.
\end{align}
This gives an orthonormal basis for the stationary, axisymmetric, nonsingular vector potentials on the Schwarzschild horizon.  The lowest order moment, $A_{1\phi}$, coincides with the Wald solution \cite{1974PhRvD..10.1680W}.  All of these multipole moments are expelled from the horizon in the extremal limit \cite{1985MNRAS.212..899B}.

\subsection{Other Possibilities}
\label{sec:other}

Split monopole fields provide a natural way to power jets in the extremal limit.  There are two other possibilities, although they are probably less important for astrophysics.  

First, non-axisymmetric fields may continue to thread the horizon in the extremal limit.  The simplest possibility is a tilted jet.  Tilted fields evade the cylinder argument of Section \ref{sec:cylinder} because they become radiative near the horizon.  The flux through the equatorial plane remains finite because the field is rapidly oscillating. Exact solutions for tilted magnetic fields have been found and the flux through the horizon is nonzero in the extremal limit \cite{1985MNRAS.212..899B}.  Astrophysical spin-powered jets are probably aligned with the black hole spin axis, so it is not clear that this possibility is astrophysically relevant.

Another way for fields to thread the horizon in the extremal limit is to consider electrically charged black holes.  Charged black holes have a gyromagnetic ratio $\gamma = Q/M$, so a charged black hole immersed in a uniform magnetic field acquires an angular momentum.  Angular momentum induces a magnetic dipole moment at the horizon.  The induced field continues to thread the horizon in the extremal limit \cite{1980PhRvD..22.2933B}.  It rises up through the throat, rather than penetrating down into it, so it is not quenched by the Meissner effect.  (The original field is expelled in the extremal limit, which shows the Meissner effect operates even for coupled electramognetic and gravitational perturbations of extreme charged black holes \cite{1980PhRvD..22.2933B}.)  However, astrophysical black holes are electrically neutral, so this possibility probably goes unrealized in nature.

\section{Conclusions}
\label{sec:conc}

We have revisited the black hole Meissner effect, whereby a spinning black hole tends to expel magnetic fields.  We have argued that if the Meissner effect operates, then electromagnetic spin-powered jets will be quenched.  This argument has a simple explanation in the membrane paradigm.  In this picture, the black hole's rotational energy is stored on the horizon.  Fields extract the black hole's energy by torquing the horizon.  If the field does not thread the horizon, then there is no torque and it is impossible to extract the black hole's rotational energy.  A subtlety is that the membrane is described by a teleological Green's function (which reflects the global nature of the event horizon), so the ``response'' precedes the torque.  This means the field need not thread the horizon until after the energy is extracted.  The membrane paradigm makes it clear that if the Meissner effect operated, then it would prevent the field from torquing the black hole and there could be no jets.

We have explained how to understand this claim without using the membrane paradigm.  The key fact is that all negative energy geodesics eventually cross the horizon.  Our claim can be understood as a generalization of this fact from particles to magnetic fields.  A magnetic field configuration which is not on a horizon crossing trajectory cannot be a negative energy configuration.  If all stationary axisymmetric fields were prevented from crossing the horizon by the Meissner effect, then there could be neither energy extraction nor BZ jets. 

Given the importance of the Meissner effect, it is crucial to understand whether astrophysical jets can avoid it and continue to operate at high spins. One might argue that a sufficiently powerful accretion disk or conductive magnetosphere would simply overwhelm the Meissner effect and drag any field configuration onto the horizon.  We have shown that this is impossible unless the field becomes radial near the horizon.  A simple example of such a field is a split-monopole.  The jets in our simulations spontaneously develop a large split-monopole component.  So this provides a natural mechanism to power jets from rapidly rotating black holes such as Cyg X-1 and GRS 1915+105.   This addresses a long-standing concern that the black hole Meissner effect quenches jet power at high spins.

 It would be good to find a simple explanation for why simulated black hole fields tend to have a large split-monopole component.   Also, as we have noted, simulations have trouble resolving the black hole throat for $a/M\gtrsim 0.95$.  The Meissner effect is a manifestation of the throat, at least from the perspective of an observer who remains outside the black hole.  (It is not clear how to understand the Meissner effect from the perspective of an infalling observer.)  So it would be good to find a way to simulate jets in a way that resolves the throat.  Perhaps the near-horizon extremal Kerr geometry could be useful \cite{1999PhRvD..60j4030B}.

\begin{acknowledgments}

It is a pleasure to thank J. Bi{\v c}{\' a}k for correspondence.  We thank J.P. Lasota, J.C. McKinney, and Y. Takamori for comments on an early draft.  R.F.P is supported by a Pappalardo Fellowship in Physics at MIT. 

\end{acknowledgments}

\bibliography{msnew}

\end{document}